\documentclass[sigconf]{acmart}

\AtBeginDocument{%
  \providecommand\BibTeX{{%
    \normalfont B\kern-0.5em{\scshape i\kern-0.25em b}\kern-0.8em\TeX}}}

\copyrightyear{2025}
\acmYear{2025}
\setcopyright{acmlicensed}
\acmConference[WWW Companion '25]{Companion Proceedings of the ACM Web Conference 2025}{April 28-May 2, 2025}{Sydney, NSW, Australia}
\acmBooktitle{Companion Proceedings of the ACM Web Conference 2025 (WWW Companion '25), April 28-May 2, 2025, Sydney, NSW, Australia}
\acmDOI{10.1145/3701716.3715537}
\acmISBN{979-8-4007-1331-6/2025/04}

\usepackage[most]{tcolorbox}
\usepackage{multirow}
\usepackage{xcolor}
\usepackage{colortbl}
\usepackage{balance}

\usepackage{enumitem}

\newcommand{\ie}{\emph{i.e., }}

\begin{document}

\title{Leveraging LLMs for Influence Path Planning in\\ Proactive Recommendation}

\author{Mingze Wang}
\orcid{0009-0009-2047-5111}
\email{gnaweinre@mail.ustc.edu.cn}
\affiliation{%
  \institution{University of Science and Technology of China}
  \city{Hefei, Anhui}
  \country{China}
}

\author{Shuxian Bi}
\orcid{0009-0002-2399-4346}
\email{shuxianbi@mail.ustc.edu.cn}
\affiliation{%
  \institution{University of Science and Technology of China}
  \city{Hefei, Anhui}
  \country{China}
}

\author{Wenjie Wang}
\authornotemark[1]
\orcid{0000-0002-5199-1428}
\email{wenjiewang96@gmail.com}
\affiliation{%
  \institution{University of Science and Technology of China}
  \city{Hefei, Anhui}
  \country{China}
}

\author{Chongming Gao}
\authornote{Corresponding Author.}
\email{chongming.gao@gmail.com}
\orcid{0000-0002-5187-9196}
\affiliation{%
  \institution{University of Science and Technology of China}
  \city{Hefei, Anhui}
  \country{China}
}

\author{Yangyang Li}
\orcid{0000-0002-8478-3932}
\email{liyangyang@live.com}
\affiliation{%
  \institution{Academy of Cyber}
  \city{Beijing}
  \country{China}
}

\author{Fuli Feng}
\orcid{0000-0002-5828-9842}
\email{fulifeng93@gmail.com}
\affiliation{%
  \institution{University of Science and Technology of China}
  \city{Hefei, Anhui}
  \country{China}
}

\renewcommand{\shortauthors}{Mingze Wang et al.}

\begin{abstract}
    
    Recommender systems are pivotal in Internet social platforms, yet they often cater to users' historical interests, leading to critical issues like echo chambers. To broaden user horizons, proactive recommender systems aim to guide user interest to gradually like a target item beyond historical interests through an influence path,\ie a sequence of recommended items. As a representative, Influential Recommender System (IRS) designs a sequential model for influence path planning but faces issues of lacking target item inclusion and path coherence. To address the issues, we leverage the advanced planning capabilities of Large Language Models (LLMs) and propose an LLM-based Influence Path Planning (LLM-IPP) method. LLM-IPP generates coherent and effective influence paths by capturing user interest shifts and item characteristics. We introduce novel evaluation metrics and user simulators to benchmark LLM-IPP against traditional methods. Our experiments demonstrate that LLM-IPP significantly enhances user acceptability and path coherence, outperforming existing approaches.


\end{abstract}

\ccsdesc[500]{Information systems~Recommender systems}

\keywords{Proactive Recommender Systems, Large Language Model, Influence Path, User Simulator}



\maketitle

\section{Introduction}

\begin{figure}[t]
  \centering
  \includegraphics[width=\linewidth]{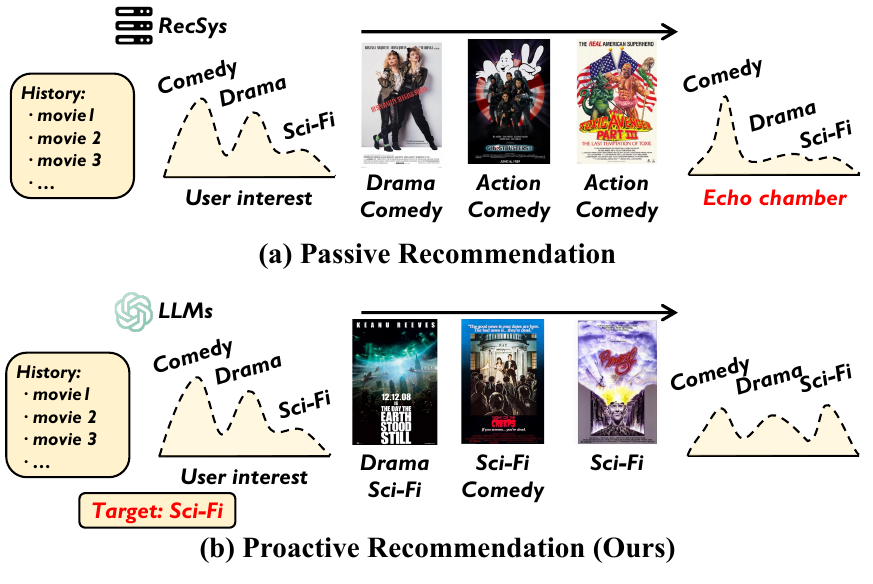}
  \caption{Illustration of Passive and Proactive Recommendation.}
  \label{img:example}
\end{figure}

Recommender system plays an important role in Internet social platforms.  
Traditionally, recommender systems infer user interest from historical behavior patterns and recommend items aligned with the user's interest~\cite{10.1145/3397271.3401125,10.1007/978-3-031-56063-7_5,RPP_TOIS,wentao2024sigir}. 
However, it will narrow the user's interest and cause the echo chamber phenomenon by continually catering to historical interest~\cite{gao2023cirs,gao2023alleviating,gao2024sprec,chen2025dlcrec}. 
To avoid this problem, researchers have proposed the task of proactive recommender system~\cite{irs}, which aims to actively guide the user interest with a sequence of items (called influence path) toward a given target item beyond historical interests, as shown in Figure~\ref{img:example}. 
The influence path gradually guides shifts in user interests through continuous recommendation exposure, facilitating the discovery of new interests. 
As such, influence path planning is the most crucial step for enhancing the efficacy of proactive recommender systems. 

The most pertinent work is Influential Recommender System (IRS)~\cite{irs}, which frames proactive recommendation as a sequential recommendation problem and constructs the influence path using a Transformer-based sequential model~\cite{10.5555/3295222.3295349}. 
Such a sequential model learns the interest-shifting patterns from users' historical data and encodes the target item's characteristics to generate the influence path. 
However, IRS encounters several critical issues: 1) There is a low probability of target item inclusion in the influence path, with a success rate below 30\% on MovieLens-1M (refer to Table~\ref{tab:certain dataset});
and 2) the adjacent items in the path are not sufficiently coherent, hindering the effective guidance of user interest. 
Moreover, directly incorporating the target item into the influence path to address the first issue will further diminish the path's coherence.

In light of these, the key objectives of influence path planning are: 1) incorporating the target item at the end of the path; 2) improving path coherence for guiding smoothness and effectiveness by ensuring adjacent items in the path are relevant; and 3) using items that users are likely to enjoy for guidance, improving user acceptability and keeping users engaged in the interaction session. 

Fortunately, Large Language Models (LLMs) possess powerful instruction-following and planning abilities~\cite{wentao2024sigir,10.1145/3626772.3657762,10.1145/3626772.3657782}, demonstrating great potential for generating a coherent influence path that includes target items. 
In this light, we propose an LLM-based Influence Path Planning (LLM-IPP) method, which elaborately prompts LLMs to capture user interest shifts and generate coherent and effective influence paths through prompt engineering techniques. 
We further design automatic evaluation metrics to compare the performance of LLM-IPP with multiple traditional influence path planning methods. 
Moreover, we present various user simulators based on LLMs and traditional recommender models to evaluate the coherence of the influence path and user acceptability towards the influence path. Empirical results validate the superiority of LLM-IPP in terms of
path coherence and user acceptability. We release our code and data at \url{https://github.com/gnaWeinrE/LLM-IPP}.

The contributions of this paper are:

\begin{itemize}[leftmargin=*]
    \item To our knowledge, this is the first attempt to integrate LLMs into proactive recommendation and we design a simple yet effective method for LLMs to achieve influence path planning.
    
    \item To evaluate the performance of LLM-IPP, we design diverse user simulators and metrics to assess user acceptability and the coherence of the influence path. 
    
    \item Extensive experiments validate that LLM-IPP significantly surpasses traditional baselines with stronger user acceptability and path coherence. 
\end{itemize}

\begin{table*}[h]
    \setlength {\tabcolsep} {1.2pt}
    \centering
    \caption{Results in MovieLens-1M and Last.FM datasets. We implement five LLM-based user simulators: GPT-3.5, GPT-4, Gemini, Llama2 7b, and Llama3 8b, and one traditional recommender system SASRec to evaluate the acceptability and coherence score. The bold number indicates the best performance, and the underlined number indicates the second-best performance. }
    \begin{tabular}{c|c|cccccc|cccccc|cc|c}
        \toprule
        &&  \multicolumn{6}{c|}{Acceptability}&  \multicolumn{6}{c|}{Coherence}&   \multicolumn{2}{c|}{SASRec}&\\
        &Method&  GPT-3.5&  GPT-4&  Gemini&  Llama2& Llama3&SASRec& GPT-3.5& GPT-4& Gemini& Llama2& Llama3&OpenAI&    IoI&IoR&SR\\
        \midrule
        \multirow{6}{*}{\begin{tabular}[c]{@{}c@{}}MovieLens\\1M\end{tabular}}&POP&  0.409&  0.576&  0.53&   0.601& 0.591&\textbf{400.5}& 0.483& 0.553& 0.474& 0.530& 0.453&\textbf{0.842}&0.801&193.9&0.008\\
        &IRS&  0.301&  0.460&  0.42&   0.564& 0.597
&973.3& 0.376& 0.457& 0.489& 0.556& 0.475
&0.800&    2.680&651.9&0.268\\
 &IRS-fix& 0.284& 0.438& 0.346& 0.575&  0.580
&1010.9& 0.365& 0.443& 0.389& 0.540&  0.458
&0.799&  2.680&651.9&1
\\
        &GPT-3.5&  \textbf{0.629}&  \textbf{0.742}&  \textbf{0.690}&  0.622& \underline{0.710}
&\underline{736.7}&\textbf{ 0.682}& 0.649& \textbf{0.653}& \textbf{0.590}& \textbf{0.631}
&\textbf{0.842}&    5.409
&1192.9&1
\\
        &GPT-CoT&  \underline{0.610}&  \underline{0.698}&  \underline{0.679}&   \underline{0.623}& 0.705
&788.9& 0.662& \underline{0.686}& \underline{0.652}& \underline{0.579}& \underline{0.625}
&\underline{0.841}&    \underline{5.450}&\underline{1215.6}&1
\\
        &GPT-ToT&  0.605&  0.682&  0.666&  \textbf{0.635}& \textbf{0.718}&796.3& \underline{0.673}& \textbf{0.701}& 0.630& \underline{0.579}& 0.617&0.839&    \textbf{5.596}&\textbf{1240.5}&1\\
        \midrule
        \midrule
        \multirow{6}{*}{Last.FM}      &POP&  0.334&  0.497&  0.414&   0.526& 0.521
&2147.5& 0.609& 0.552& 0.518& 0.546& 0.456
&\underline{0.810}&    0.914&619.8&0\\
        &IRS&  0.406&  0.618&  0.556&   0.567& 0.525
&\textbf{1255.4}& 0.520& 0.563& 0.492& 0.506& 0.437
&\textbf{0.811}&    0.723&516.1&0.142\\
 &IRS-fix& 0.441& 0.563& 0.483& 0.560&  0.524
&\underline{1627.0}& 0.541& 0.533& 0.536& 0.525&  0.429
&0.809&  0.723&516.1&1\\
        &GPT-3.5&  0.532&  0.539&  0.542&  0.582& \underline{0.660}
&4145.6& \textbf{0.729}&\textbf{0.803}& \underline{0.652}& \textbf{0.591}& \underline{0.667}
&0.802&    6.159&1447.5&1\\
        &GPT-CoT&  \underline{0.545}&\underline{0.558}&\underline{0.584}&\underline{0.590}& \textbf{0.672}
&4002.8& \underline{0.725}& \underline{0.776}& \textbf{0.700}& \underline{0.589}& \textbf{0.670}
&0.809&    \textbf{6.354}&\underline{1544.5}&1\\
        &GPT-ToT&  \textbf{0.547}&  \textbf{0.634}&  \textbf{0.618}&   \textbf{0.592}& 0.652&3824.0& 0.718& 0.731& 0.594& 0.585& 0.657&0.803&    \underline{6.291}&\textbf{1546.5}&1\\
        \bottomrule
    \end{tabular}
    \label{tab:certain dataset}
\end{table*}

\section{Method}

In this study, we propose an innovative method that leverages the instruction-following and path-planning capabilities of LLMs to generate influence paths. Our design involves the inclusion of target items, maintaining coherence, and incorporating items of interest to the user into the instructions, prompting the LLMs to utilize their world knowledge to create paths that meet the conditions. To further enhance the performance, we employ prompt engineering techniques to improve the interest-guiding ability of LLM-IPP.

\subsection{Influence Path Generation}

LLM-IPP leverages the user's demographic features and historical data sequence to prompt LLMs to generate the influence path. For example, in the MovieLens-1M dataset, the prompt is as shown on the right.

In the prompt, we set several constraints for the LLMs. (1) The LLMs are required to connect the user interest with the target item, implying that the influence path must include the target item. (2) Direct recommendation of the target item to the user is prohibited; therefore, a minimum length for the influence path is essential. (3) Adjacent items should exhibit a strong relation to instructing the LLMs to generate an influence path with high coherence. 
Additionally, to make the recommendation more reasonable, the generated items should meet the dataset-specific limitations. For example, for the MovieLens-1M dataset, the items are all movies before 2001, thus we need to constrain the output space in the prompt. For the Last.FM dataset, the items are musical artists before 2011.

\begin{tcolorbox}[title = {Prompt for MovieLens-1M dataset}]
\textbf{(Optional: ToT)} Imagine three different experts answering this question. All experts will write down 1 step of their thinking and then share it with the group. Then all experts will go on to the next step, etc. And show me the result in the end. The question is:\\
\vspace{-10pt}
\tcbline
You are a recommender system. Given the user profile and historical data, analyze the user's interests. Your task is to add no less than five movies between the last movie of historical data and the target movie to connect them. Any adjacent movies should have a strong relation (share the same genre, theme, title, etc.) with each other and make sure the movies are not included in the historical data. The release date of the movies should be before 2001. \\
\vspace{-10pt}
\tcbline
\textbf{(Optional: CoT)} Think step by step and make sure. \\
\vspace{-10pt}
\tcbline
<User's demographic feature>: Gender/Age/Occupation \\
<Historical data>: sequence of <Movie title, Movie genre>\\
<Target item>: <Movie title, Movie genre>\\
\vspace{-10pt}
\end{tcolorbox}
  

We also discovered that the commonly used prompt engineering techniques, such as ToT~\cite{long2023large} and CoT~\cite{wei2023chainofthought}, are effective on LLM-IPP. Only minor modifications to these techniques are required to enhance its performance. For example, CoT needs the addition of \textbf{``Think step by step and make sure.''} at the end of the prompt. For ToT, we based on Hubert's methodology~\cite{tree-of-thought-prompting}, employing a straightforward prompting engineering approach to integrate the principal concept of ToT into LLM-IPP. 

\section{Experiment}

In this section, we conduct experiments on two real-world datasets and design simulators with corresponding metrics to evaluate the influence paths. 
LLMs are used to simulate users to provide feedback and leverage LLM-based metrics to evaluate the performance of influence paths. 
Traditional recommender models can also be developed as user simulators, usually implemented by the Transformer model and word embeddings~\cite{8594844} to assess the influence paths. 

\textbf{Certain-domain Dataset.} We conduct the experiment on two real-world datasets, MovieLens-1M~\footnote{\url{https://grouplens.org/datasets/movielens/1m/}.} and Last.FM~\footnote{\url{https://grouplens.org/datasets/hetrec-2011/}.} dataset. MovieLens-1M dataset contains 6,038 users and 3,883 movies, totaling 5,084 user history item sequences, excluding sequences shorter than 20. Last.FM dataset contains 1,892 users and 17,632 musical artists, totaling 1,297 user history item sequences.
In this work, the target items are randomly sampled from all items. For training, we separate the dataset into training and testing sets accounting for 20\% and 80\% respectively. 
The traditional baselines are trained on the training sets and evaluated on the testing sets. 
LLM-IPP is directly evaluated on the testing sets without training since it is a zero-shot method.

\subsection{Metrics}

To thoroughly evaluate the methods, we employ both LLM-based and traditional metrics. LLM-based metrics utilize independent LLMs to represent the users to provide feedback and evaluate the performance of influence paths. 
Traditional metrics are based on the simulator implemented by the Transformer model and word embeddings to calculate the performance of influence paths.

\subsubsection{Traditional Recommender System Simulator}

We train an independent next-item recommender system that works as a user simulator for evaluation~\cite{ipg, irs}. In this study, SASRec~\cite{8594844} is employed as the base model, which utilizes a self-attention-based model that captures both long-term and short-term semantics. SASRec learns user behaviors from the historical item sequence and outputs the probability distribution of items that represent the user's current preference. In addition, by incorporating word embedding methods and probability theory, we employ the following metrics to evaluate the performance of the influence path.

\textbf{Success rate (SR)} measures the ratio of the generated influence path that contains the target item.

\textbf{Increase of Interest (IoI)~\cite{irs}} measures the change in the user’s interest in the target item after being persuaded through the influence path.
\begin{equation}
\textrm{IoI} = \frac{1}{|U|} \sum_{u=1}^{|U|} (\log{P(i_t^u|seq^u \oplus seq_t^u)} - \log{P(i_t^u|seq^u)}),
\end{equation}
where $\oplus$ denotes the concatenation of two sequences; $seq^u$ is the history item sequence of user $u$; $i_t^u$ is the target item set for user $u$; $seq_t^u$ is the generated influence path; $P(i|seq)$ is the probability of item $i$ when input the sequence $seq$ into the SASRec model.

\textbf{Increase of Rank (IoR)~\cite{irs}} measures how much the target item’s ranking improves after persuasion via the influence path.
\begin{equation}
\textrm{IoR} = -\frac{1}{|U|} \sum_{u=1}^{|U|} (R(i_t^u|seq^u \oplus seq_t^u) - R(i_t^u|seq^u)),
\end{equation}
where $R(i|seq)$ is the rank of item $i$ when input the sequence $seq$ into the SASRec model.

\textbf{Acceptability (SASRec)} measures user acceptability by calculating the mean ranking of the items in the influence path.
\begin{equation}
\textrm{Acceptability} = \frac{1}{|U|}  \sum_{u=1}^{|U|} \frac{1}{|seq^u_t|} \sum_{i \in seq_t^u} R(i|seq^u).
\end{equation}
\textbf{Coherence (OpenAI)} measures the relevance of each of two adjacent items in the influence path by calculating the mean cosine similarity of OpenAI Embeddings~\cite{openai2022embeddings}.
\begin{equation}
\textrm{Coherence} = \frac{1}{|U|} \sum_{u=1}^{|U|} \frac{1}{|seq^u_t|} \sum_{i_j,i_{j+1} \in seq^u_t} \textrm{cos\_sim}(i_j, i_{j+1}).
\end{equation}

\subsubsection{LLM-based Simulator}

Except for the traditional recommender system simulator, to appropriately evaluate the LLM-based method, we introduce the LLM-based Simulator. Traditional metrics are incapable of revealing the latent relationships between the items, unlike LLMs, due to the lack of world knowledge about the recommended items. 
The most convincing evaluation method is implementing LLM-IPP on an online social platform and collecting real user feedback, which is significantly expensive. 
Fortunately, many recent studies on LLMs imply that the LLMs can be an appropriate assessment tool, even for recommender systems~\cite{tokutake2024large, zhang2024largelanguagemodelsevaluators}. In this paper, we implement commonly used LLMs such as GPT, Gemini, and Llama to provide scores for the influence paths on each metric.

\textbf{Acceptability (LLMs)} measures the user’s acceptance of the influence path and interest in the target item via LLMs.

\begin{tcolorbox}[title = {Prompt for measuring user acceptability}]
Given the user profile and the historical data, analyze the user's interest. Based on this information, would the user be interested in the movies in the influence path step by step? Answer with a probability for each movie between 0 and 1, where 0 means ``definitely not interested'' and 1 means ``definitely interested''. Please explain the reason for each score. If uncertain, make your best guess. \\
\vspace{-10pt}
\tcbline
<User's demographic feature>: Gender/Age/Occupation \\
<historical data>: sequence of <Movie title, Movie genre>\\
  <influence path>: sequence of <Movie title, Movie genre> \\
\vspace{-10pt}
\end{tcolorbox}

  

\textbf{Coherence (LLMs)} measures the mean relevance of each of two adjacent items by asking the LLMs.

\begin{tcolorbox}[title = {Prompt for measuring path coherence}]
Given the influence path in the format of A,B,C..., what's the relevance of each adjacent item? Answer with a score between -1 and 1, where -1 means ``definitely not related'' and 1 means ``definitely related''. Please explain the reason for each score. If uncertain, make your best guess\\
\vspace{-10pt}
\tcbline
<influence path>: sequence of <Movie title, Movie genre>\\
\vspace{-10pt}
\end{tcolorbox}

  

\subsection{Baselines \& Proposed Methods}

We implement three baselines and three LLM-IPP methods.

\textbf{IRS~\cite{irs}} is the SOTA traditional proactive recommender system. Improve the performance by designing a Transformer-based sequential model.

\textbf{IRS-fix} fixes the baseline's low success rate. We append the target item to the IRS-generated influence path if it doesn't contain the target item to reach a 100\% success rate.

\textbf{POP} sorts all the items by occurrence and recommends the most popular items to the user.

\textbf{LLM-IPP (GPT-3.5)} is the base version of LLM-IPP.

\textbf{LLM-IPP (GPT-CoT)} enables complex reasoning capabilities via intermediate reasoning steps.

\textbf{LLM-IPP (GPT-ToT)} introduces a framework that generalizes over CoT prompting and encourages exploration over thoughts that serve as intermediate steps for general problem-solving with language models.


\subsection{Analysis}

Table~\ref{tab:certain dataset} presents the results of the methods evaluated in the MovieLens-1M dataset and the Last.FM dataset. From the table, it is evident that LLM-based methods achieve higher scores on most of the metrics than other baselines. For the traditional metrics, all the LLM-based methods significantly outperform traditional baselines in $IoI$ and $IoR$, demonstrating that LLM-IPP effectively improves the user's interest towards the target item. 
For the LLM-based metrics, an interesting trade-off between the influence path's acceptability and coherence is observed, especially in Last.FM dataset. This phenomenon is possible because when the items in the influence path are closer to the user's interests and achieve a higher acceptability score, it will result in the items not being evenly distributed along the influence path and consequently cause a lower coherence score. 

\textbf{Case study.} In addition to the quantitative study, we conduct a real case to demonstrate the effectiveness of LLM-IPP. As shown in Table~\ref{tab:example}, \textit{The Great Dictator} is selected as the target item. At first, LLM-IPP recommends movies in the genre of ``Drama''. After a few attempts, \textit{Being There}, with the genres of both ``Drama'' and ``Comedy,'' is recommended. Then finally to the target item.

However, LLM-IPP reaches further beyond the movie names and genres, which are the only details provided by the MovieLens-1M dataset. With LLM's great ability in natural language understanding and world knowledge, LLM-IPP knows the movies' cast, director, content, and other movie attributes. In this example, based on the user's other viewing history, LLM-IPP considers this user is likely to enjoy movies with the genres of ``War'' and ``Drama'' and enjoys movies with strong narratives and suspenseful plots. According to the recommendation reasons generated by LLM-IPP, \textit{Path of Glory} and \textit{Dr. Strangelove} are both directed by Stanley Kubrick and have a war theme. \textit{Network} and \textit{Being There} are both about media critique and dark comedy. These understandings about the movies help to build a more practical influence path and reach higher acceptability and coherence scores. The results affirm the effectiveness of LLM-IPP in smoothly leading the user preference to the target items.

\begin{table}[t]
\caption{An example of an influence path.}
\label{tab:my_table}
\begin{tabular}{>{\raggedright\arraybackslash}p{3.9cm}>{\raggedright\arraybackslash}p{3.9cm}}
\toprule
\textbf{Name}&\textbf{Genre}\\
\hline
\multicolumn{2}{l}{\textbf{The last movie in the viewing history}}\\
Frequency (2000)&Drama, Thriller\\
\hline
\textbf{Influence path}&\\
Paths of Glory (1957)&Drama, War\\
Dr. Strangelove (1964)&Comedy, War\\
Network (1976)&Drama\\
Being There (1979)&Comedy, Drama\\
To Be or Not to Be (1942)&Comedy, Romance, War\\
\hline
The Great Dictator (1940)&Comedy\\
\bottomrule
\end{tabular}
\label{tab:example}
\end{table}

\textbf{User Study.} We conduct a double-blind human test to evaluate the performance of LLM-IPP. We recruit three volunteers who are experienced in the field of information retrieval. For the evaluation, we create a total of 100 examples. For each example, we provide the volunteers with the user’s historical data along with two influence paths generated by the baseline model (IRS) and LLM-IPP (GPT-ToT), respectively. The volunteers are asked to determine the superior influence path. The preference rates for LLM-IPP among the three volunteers were 66.7\%, 63.6\%, and 84.2\%. The average Cohen’s kappa score was 0.63, indicating a high level of agreement among the volunteers.

\section{Conclusion \& Discussion}
This paper proposed an LLM-based Influence Path Planning (LLM-IPP) method for proactive recommendation, which can generate the influence path with high coherence and user acceptability. 
We explored several prompting techniques to identify the optimal prompt to instruct LLMs for influence path planning. Additionally, to evaluate LLM-IPP, we implemented various LLM-based and traditional simulators and metrics to ensure a comprehensive assessment of user acceptability and path coherence. Experimental results demonstrate that the LLM-IPP outperforms traditional proactive recommender systems in terms of path coherence and user acceptability.

As the first work to apply LLMs to proactive recommender systems, we leave many promising research directions: 
(1) While extensive user studies are essential, it is hard for participants to simulate the interest shifts across multi-round recommendations. How to design an effective user evaluation for proactive recommendation is a future direction. 
(2) It is feasible to fine-tune LLM-IPP with real-world data, injecting real-world user interest shifting patterns into LLMs for influence path planning. 
(3) In addition to prompt engineering, other optimization techniques related to LLMs can be applied to proactive recommendation systems. For instance, employing LLM-based agents that use memory mechanisms to improve user satisfaction or fine-tuning LLMs to better align them with proactive recommendation tasks may enhance their performance.

\section*{Acknowledgements}
This work is supported by the National Natural Science Foundation of China (62402470,62272437,U21B2026), the Fundamental Research Funds for the Central Universities of China (WK2100000053, PA2024GDSK0107), Anhui Provincial Natural Science Foundation (2408085QF189), and the Postdoctoral Fellowship Program of CPSF (GZC20241643). This research is supported by the advanced computing resources provided by the Supercomputing Center of the USTC.

\bibliographystyle{ACM-Reference-Format}
\balance
\bibliography{main}

\appendix


\end{document}